\documentclass[9pt, sigplan,authorversion]{acmart}
\usepackage{graphicx}

\usepackage{svg}
\usepackage{adjustbox}
\usepackage{csquotes}
\usepackage{amsmath}
\usepackage{subcaption}

\usepackage{listings}
\usepackage{xcolor}

\colorlet{punct}{red!60!black}
\definecolor{background}{HTML}{EEEEEE}
\definecolor{delim}{RGB}{20,105,176}
\colorlet{numb}{magenta!60!black}

\lstdefinelanguage{json}{
    basicstyle=\normalfont\ttfamily,
    numbers=left,
    numberstyle=\scriptsize,
    stepnumber=1,
    numbersep=8pt,
    showstringspaces=false,
    breaklines=true,
    frame=lines,
    literate=
     *{0}{{{\color{numb}0}}}{1}
      {1}{{{\color{numb}1}}}{1}
      {2}{{{\color{numb}2}}}{1}
      {3}{{{\color{numb}3}}}{1}
      {4}{{{\color{numb}4}}}{1}
      {5}{{{\color{numb}5}}}{1}
      {6}{{{\color{numb}6}}}{1}
      {7}{{{\color{numb}7}}}{1}
      {8}{{{\color{numb}8}}}{1}
      {9}{{{\color{numb}9}}}{1}
      {:}{{{\color{punct}{:}}}}{1}
      {,}{{{\color{punct}{,}}}}{1}
      {\{}{{{\color{delim}{\{}}}}{1}
      {\}}{{{\color{delim}{\}}}}}{1}
      {[}{{{\color{delim}{[}}}}{1}
      {]}{{{\color{delim}{]}}}}{1},
}

\newtheorem{mydef}{Definition}

\copyrightyear{2023}
\acmYear{2023}
\setcopyright{acmlicensed}\acmConference[MiddleWEdge '23]{2nd International Workshop on Middleware for the Edge}{December 11, 2023}{Bologna, Italy}
\acmBooktitle{2nd International Workshop on Middleware for the Edge (MiddleWEdge '23), December 11, 2023, Bologna, Italy}
\acmPrice{15.00}
\acmDOI{10.1145/3630180.3631202}
\acmISBN{979-8-4007-0451-2/23/12}

\begin{document}

\title[Offloading Real-Time Tasks under Consideration of Networking Uncertainties]{Offloading Real-Time Tasks in IIoT Environments under Consideration of Networking Uncertainties}

\author{Ilja Behnke}
\orcid{0000-0002-2437-8994}
\email{i.behnke@tu-berlin.de}
\affiliation{%
    \institution{TU Berlin}
    \city{Berlin}
    \country{Germany}
}

\author{Philipp Wiesner}
\orcid{0000-0001-5352-7525}
\email{wiesner@tu-berlin.de}
\affiliation{%
    \institution{TU Berlin}
    \city{Berlin}
    \country{Germany}
}

\author{Paul Voelker}
\orcid{0009-0001-9366-576X}
\email{mail@paulvoelker.de}
\affiliation{%
    \institution{TU Berlin}
    \city{Berlin}
    \country{Germany}
}

\author{Odej Kao}
\orcid{0000-0001-6454-6799}
\email{odej.kao@tu-berlin.de}
\affiliation{%
    \institution{TU Berlin}
    \city{Berlin}
    \country{Germany}
}

\begin{abstract}
Offloading is a popular way to overcome the resource and power constraints of networked embedded devices, which are increasingly found in industrial environments.
It involves moving resource-intensive computational tasks to a more powerful device on the network, often in close proximity to enable wireless communication.
However, many Industrial Internet of Things (IIoT) applications have real-time constraints.
Offloading such tasks over a wireless network with latency uncertainties poses new challenges.

In this paper, we aim to better understand these challenges by proposing a system architecture and scheduler for real-time task offloading in wireless IIoT environments.
Based on a prototype, we then evaluate different system configurations and discuss their trade-offs and implications. Our design showed to prevent deadline misses under high load and network uncertainties and was able to outperform a reference scheduler in terms of successful task throughput. Under heavy task load, where the reference scheduler had a success rate of 5\%, our design achieved a success rate of 60\%.
\end{abstract}




\maketitle         
\renewcommand{\shortauthors}{Behnke et al.}

\section{Introduction}

The Industrial Internet of Things (IIoT) aims to connect embedded devices in industrial environments to enable remote control, mobile computing, monitoring, and collaborative problem solving.
Examples can be found in the cyber-physical systems used in smart factories, logistics, and autonomous vehicles~\cite{bock2021performance,barzegaran_fogification_2020}, where devices often operate under hard real-time constraints.
While real-time computing in automation and embedded systems is well researched and deployed, its combination with packet-based wireless IP networking presents new challenges and opportunities. 

Devices used in IIoT scenarios typically have resource and power constraints that limit the type and amount of local computation.
For example, resource-intensive tasks such as object recognition in video may require specialized hardware and more power than an embedded device can reasonably provide. Another example for this can be found in close-proximity model training, where sensor data is accumulated and processed on edge servers~\cite{becker2021local}.
A common way to deal with these limitations in practice is to offload individual tasks to more powerful computers in the local area~\cite{iiot_edge,iiot_offloading_energy,iiot_offloading_ai}.
However, for tasks with real-time requirements, this can be very challenging due to unpredictable latencies in wireless networks, especially when considering mobile devices.

Scheduling real-time tasks in distributed systems has received increasing attention~\cite{roy2020contention,yi2020task,behnke2020interrupting}, but related work generally assumes very reliable and predictable communication.
However, especially with wireless connections, the exact prediction of the expected latency is difficult.
Various efforts have been made to develop methods to guarantee end-to-end delay in wireless networks, but some form of uncertainty always remains~\cite{e2e_delay_1,e2e_delay_2,e2e_delay_3,behnke2023towards}.
In addition, efficiently deploying workloads in a distributed system is difficult, especially when these workloads occur sporadically~\cite{scheduling_difficult}.
In this work, we aim to better understand the tradeoffs and implications of offloading real-time tasks to local edge resources in IIoT environments connected by wireless networks.
To this end, we make the following \emph{contributions}:

\begin{itemize}
    \item We present a system architecture for real-time task offloading in IIoT environments
    \item We discuss and implement a distributed real-time scheduler incorporating network latency uncertainties
    \item We evaluate relevant metrics using a prototype of our system within a virtual environment
\end{itemize}

The reminder of this paper is structured as follows.
Section~\ref{sec:related} surveys the related work.
Section~\ref{sec:sysarch} explains our system architecture.
Section~\ref{sec:scheduler} proposes a scheduler to perform latency-aware offloading.
Section~\ref{sec:evaluation} evaluates different system configurations to quantify trade-offs between performance and reliability.
Section~\ref{sec:conclusion} concludes the paper.
\section{Related Work}
\label{sec:related}
In this section, we briefly present similar work in the area of real-time task offloading.

Chen et al.~\cite{iiot_offloading_energy} use fog computing for energy-optimal dynamic computation offloading in IIoT environments.
Computation time, energy, and resource utilization are considered and optimized. Real-time guarantees are not taken into account. Similarly, Elashri and Azim~\cite{elashri_energy-efficient_2020} use offloading of real-time tasks to cloud and fog resources for energy efficiency. They propose two algorithms for making an efficient offloading decision for soft and weakly hard (firm) real-time applications while guaranteeing task schedulability.

Ma et al.~\cite{ma_reliability_2022} identify a trade-off problem between reliability and latency in IIoT applications.
In their work, they propose an offloading framework for visual applications that considers both performance metrics.
A scheme is presented that uses and maximizes a utility function that balances reliability and latency.
Shukla and Munir~\cite{shukla_efficient_2017} present an offloading architecture for IoT devices. Computers on the same local network as the IoT device process data-intensive tasks while meeting their real-time deadlines.

Liu et al. address the problem of offloading data- and compute-intensive computations for vehicular networks~\cite{liu_real-time_2020}. Using a fog/cloud architecture, tasks are classified and distributed according to latency requirements. Roadside units as well as other vehicles can act as compute nodes. They formulate a real-time task offloading model that maximizes the task service ratio and propose a real-time task offloading algorithm that works cooperatively.

Hong et al. present a cross-layer cooperative offloading strategy for multi-hop scenarios~\cite{hong_multi-hop_2019}. IIoT devices, local edge servers, as well as cloud resources are considered. QoS-Aware computation offloading in a distributed manner with game theory approaches.

In the following sections, we complement these works by presenting a simple architecture and real-time scheduler for IIoT environments, focusing on the incorporation of network delays as well as the tradeoff between predictability and throughput.

\section{System Architecture}
\label{sec:sysarch}

\begin{figure}
    \centering
    \includegraphics[width=.9\columnwidth]{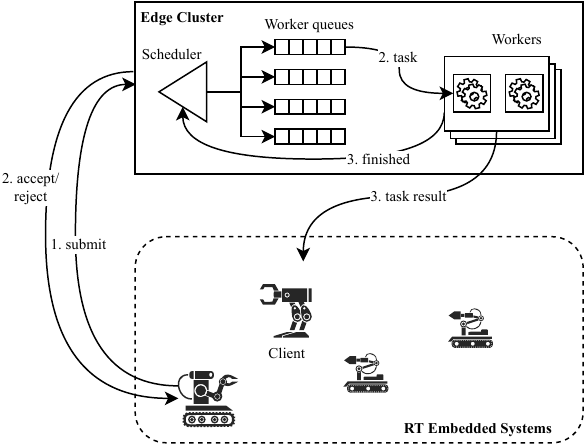}
    \caption{System Architecture Overview}
    \label{fig:overview}
\end{figure}

The goal of this work is a system design for real-time task offloading for mobile embedded systems and the development and evaluation of a scheduler for this purpose. Figure~\ref{fig:overview} shows an overview of the design. Potentially mobile machines (clients) submit tasks with deadlines to the scheduler, which checks time feasibility and -- if accepted -- schedules the tasks to worker queues. The workers run on the same compute cluster as the scheduler and can be preempted by it. Each worker computes one task at a time and returns the result directly to the source client while signaling task completion to the scheduler.

The considered environments contain a scalable number of (semi-) autonomous devices that reside and potentially move in a local area spanned by an IP network. Based on these settings, the following task and scheduler properties can be derived.

\begin{itemize}
	\item Tasks arrive \textit{sporadic}: The devices that submit the tasks are fully or semi autonomous and therefore the arrival of tasks happens during runtime, irregular and not predictable for the scheduler.
	\item The scheduling must therefore happen \textit{online} during system run-time, with \textit{dynamic priorities}.
	\item The clients might not be fully independent in the overall setting, however, their interaction is assumed to happen outside of the task context. Therefore the tasks are independent and \textit{no precedence constraints} need to be considered.
	\item Tasks have \textit{hard deadlines}, that are known at the time the task is submitted. 
	\item Tasks are \textit{fully preemptable}
	\item Finally, the computation model is that of a \textit{homogeneous multiprocessor}. However, with the major difference of distributed processing entities (workers) and task dispatch via local network.
\end{itemize}

The following subsections describe the design of the involved entities. Design decisions and the operation of the scheduler are explained in detail in Section~\ref{sec:scheduler}.

\subsection{Tasks}
\label{sec:tasks}

The \textit{task} is the central data structure that holds all the necessary information about the workload that a client wants to offload.
It is modeled after real-time operating system processes and is shared and sent between the different entities in the system.
In an IIoT environment such as a smart factory, we assume that all entities and workloads are known before the system is set up.
Therefore, clients can assume that the programs or binaries necessary to compute all the workloads that a client wants to offload are present on the workers.
This makes a task submission much like a \textit{Remote Procedure Call (RPC)}.
A task is generated initially by the client, who sends it to the scheduler, which in turn eventually sends it to a worker.
It is serialized and deserialized for transportation over the network, but all entities work on the same shared Task data structure.

\begin{mydef}
\label{def:1}
Let $\mathcal{T}$ be a task offloaded by a client, scheduled, and computed by a worker. For the purpose of this work it is characterized by the tuple $$(C, T_d, t_r, t_\text{cs}, t_w, t_e, \mathcal{P})$$
\begin{itemize}
\item $C$ the source client. 
\item $T_d$ the absolute deadline.
\item $t_r$ the initial relative deadline.
\item $t_\text{cs}$ the connection setup time.
\item $t_w$ the worst-case execution time.
\item $t_e$ the elapsed execution time.
\item $\mathcal{P}$ the set of parameters.  
\end{itemize}
\end{mydef}

The client $C$ specifies its own id and its IP address and port on which it will listen for the connection that the worker will establish.
$t_r$ is the relative deadline at the time the client creates the task and sends it to the scheduler.
This value, along with the connection setup time $t_\text{cs}$ measured by the client, is used by the scheduler to calculate network latency.
The duration $t_e$ is set and used by the scheduler to keep track of the execution time when a task is computed by a worker and possibly preempted.
The set of parameters $\mathcal{P}$ contains the command line arguments and payload data that the worker passes to the actual executable.

\paragraph{Task Dispatch}
There are two sensible design choices for when to dispatch consecutively scheduled tasks to the assigned workers:

\begin{enumerate}
    \item Dispatch a task to the worker as soon as it is assigned to it, and let the worker maintain an internal task queue.
    \item Dispatch the next task only after the previous one is finished.
\end{enumerate}

The first option has the advantage that consecutively scheduled tasks are present on the worker and can be computed with minimal interruption in between.
On the other hand, the scheduler sends lower priority tasks to the worker to add to its queue while it is working on the highest priority task.
The second option interrupts the worker only when a task with a higher priority than the one the worker is currently processing arrives.
For the remainder of this work, the second approach is used to ensure minimal disruption and maximum predictability of the execution time of the running task. 
We assume that the connection between the workers and the scheduler is wired and very fast (compared to the wireless connection of the clients), so that task dispatching causes little idle time for the workers.
In a scenario where tasks come with large input payloads, such as multimedia, the first approach may be more appropriate.

\paragraph{Task Payloads}
Tied to the task dispatch approaches is the question of what path the task payloads take.
Task payloads refer to the input parameters that the client sends with the task and the task result that is sent back to the client.
There are three possible paths for the input parameters:

\begin{enumerate}
    \item The client passes only the task metadata to the scheduler, which then communicates the assigned worker back to the client, so it can send the input parameters directly to the worker.
    \item The client sends the task together with its input parameters to the scheduler, which eventually passes them on to the assigned worker.
    \item The client uploads the input parameters to a shared storage outside the scheduler that all workers have access to.
\end{enumerate}

Based on the task dispatch handling decision, the second option is chosen for task input parameter handling.
In a scenario where the task input parameters consist of large amounts of data, such as multimedia, the scheduler might become a bottleneck if it has to handle the distribution of this data.
In this case, one of the other approaches might be more appropriate.
The task result payload is only relevant to the client. Therefore, the worker sends it directly to the client.

\subsection{Clients}
\textit{Clients} are the entities that generate \textit{tasks} they want to offload.
They are designed as a software library that provides an interface for submitting tasks.
Client machines are mobile embedded devices that perform some kind of physical action and are connected to the rest of the network via a wireless link. They implement fallback behavior in case a task is rejected by the scheduler.

\paragraph{Task configuration}

The clients are responsible to set the following properties of offloaded tasks as defined in Section~\ref{sec:tasks}. 

\begin{itemize}
    \item Execution time
    \item Deadline
    \item Result payload size
\end{itemize}

\paragraph{Task Rejections} \label{sec:design_client_task_rejections}
Submitted tasks are either accepted or rejected by the scheduler.
In case a task is accepted, the client will assume that the task result is available to the client before the deadline.
If it arrives after the deadline, this means a failure of the system and is to be generally avoided.
So, if the scheduler is not able to schedule the task on one of the workers without nearing the deadline by a certain uncertainty window, it rather rejects the task completely.
The client has some fallback behavior for this case, which causes less disruption than a missed deadline.

\subsection{Workers}
Workers compute the offloaded tasks distributed by the scheduler.
They can be thought of as distributed processing cores: A worker can process exactly one task at a time and is fully utilized with that execution.
Worker machines are edge resources that are physically close to the scheduler host and on the same local area network.
However, it listens for incoming tasks even while processing one, since tasks need to be preemptable in a setting with deadlines.
They ensure predictable task execution times.%

Workers connect to the scheduler at startup and listen for tasks.
Upon receiving a task, the worker begins executing the task with the parameters provided by the client, but continues to listen for incoming tasks to allow for preemption by the scheduler.
When the worker receives a task while still processing an earlier task, the earlier task gets preempted until the new task is completed.

\paragraph{Worker Operation} \label{sec:impl_worker_functionality}

After the worker successfully connects to the scheduler, it enters the main loop, where it waits for tasks to be sent by the scheduler through the open connection.
When a task is received, a new thread is started to handle that task while the main loop continues to listen for messages from the scheduler.

\paragraph{Worker Preparation}

The task handler running in the new thread then starts to prepare the task computation. 
The TCP 3-way handshake takes a non-negligible amount of time when performed on a connection with a latency in the double-digit millisecond range~\cite{pei2016wifi}.
For this reason, the worker starts establishing the connection to the client immediately after receiving a task, in parallel with the actual task computation.
The worker sends the computation results of a task directly to the client.
This ensures that the task result can be sent with the shortest possible delay once the task has finished computing.

\paragraph{Worker Computation}
The worker computes the tasks by starting the task executable as a separate operating system process.%
Once the process is started and a process id (PID) is returned by the operating system, the running task and its process ID are pushed into the internal task queue.
The thread then sleeps until the process has exited.
If there is already a task running, the worker lets the operating system suspend the process execution and marks the process as preempted in the task queue.
Only then the new process to compute the new task is started and the thread goes to sleep until the new process has exited.

Once the process has exited, the worker reads the computation result and packages it for transport to the client. 
The worker also records the actual execution time (excluding the time spent being preempted).
The worker notifies the scheduler of having finished the task and sends the task result to the client.
However, this needs an established connection to the client, which was started in parallel with the task computation.
If the connection setup takes longer than the task computation, the worker must wait until the open connection is available.
Finally, if some task was preempted, the halted process is resumed.

\section{The Scheduler}
\label{sec:scheduler}
This section explains the design choices and operation of the scheduler, which dispatches tasks submitted by clients to workers.
After parsing a task message into the corresponding task structure, the \textit{time to deadline} is calculated from the absolute deadline $T_d$ of the task and the current system time $T_s$.
For each incoming task, the scheduler then derives the laxity~$= T_d - T_s - (t_w - t_e)$ and density~$= (t_w - t_e) / (T_d - T_s)$ to make a scheduling decision.

\subsection{Scheduling Algorithm: Partitioned EDF}
Based on the task model and the assumed environment, only dynamic priority, preempting scheduling algorithms such as \textit{Earliest Deadline First (EDF)} and \textit{Least Laxity First (LLF)} are suitable. LLF has the major drawback of thrashing~\cite{llt}, which is exacerbated in this scenario. Since context switches are expensive, the high frequency of context switches caused by the thrashing effect is already problematic in a general-purpose operating system.
In our scenario, where each preemption not only causes a context switch on the worker, but also requires communication, the effect has an even greater impact on overall system performance.
For this reason, EDF was considered a more reasonable algorithm for our scheduler.
Yet, for the sake of completeness, there also exist enhanced versions of \textit{LLT} which aim to avoid thrashing.

In a multiprocessor environment, EDF works either with a global task queue (global EDF) or with partitioned queues for each worker (partitioned EDF).
In addition to the non-optimality of global EDF due to Dhall's effect~\cite{dhall1978real}, a global queue also makes task acceptance testing more difficult:
When a task is added to the global task queue and given its priority, it is not immediately clear which worker will eventually pick up the task, at what time, and whether this is early enough to meet the deadline.
It would be necessary to simulate separate run queues for each worker to find the next worker to become free when the new task eventually becomes the highest priority task in the global queue.
Since the scheduler must decide whether it can guarantee the timely execution of tasks as they arrive, partitioned scheduling is more appropriate.
Partitioned EDF requires an algorithm or heuristic to assign a task to a worker when multiple workers can meet the task deadline.
First-fit, worst-fit, and best-fit are the heuristics considered. Which one to use will depend on the circumstances and will be further explored in the evaluation.

\subsection{Deadline Adjustment \& Acceptance Checks}
\label{sec:adjustment}
When a new task arrives at time $t$, the scheduler adjusts the deadline set by the client to account for network latency.
In our setting, the communication delay between a client and the scheduler is roughly the same as between a worker and the client.
Therefore, the scheduler uses the delay that occurs in the client-scheduler connection to calculate the expected delay $d_\text{exp}$ that will occur when the worker sends the task result to the client:
The client provides the \textit{connection setup time} $t_\text{cs}$ and the initial time to deadline $t_r$ at the time the client generated the task in the task message. 
$d_\text{exp}=t_r-T_d-t$. 

However, in a high latency environment with short task execution times, it may take longer to establish the connection than to complete the task computation.
In this case, the difference between $t_\text{cs}$ and $t_w$ must also be considered for the delay adjustment $d_\text{adj}$:
\begin{equation*}
d_\text{adj} = \begin{cases}
d_\text{exp}+t_\text{cs}-t_w & \text{, }t_\text{cs} > t_w\\
d_\text{exp} &\text{, else}
\end{cases}
\end{equation*}
The \textit{adjusted delay} is then multiplied by the scheduler's uncertainty factor $\mathcal{U}$ to account for latency jitter.
$\mathcal{U}$ is a configuration option of the scheduler and the effect of different values is one of the subjects of the evaluation.
Finally, the resulting value is subtracted from the deadline. $T_{d\text{-new}}=T_{d\text{-old}} - \mathcal{U} \cdot d_\text{adj}$.
This adjusted deadline can now be used for scheduling and assessed to see if it leaves enough time for the task result to be sent to the client and not miss the original deadline set by the client.
A first acceptance check is then performed to ensure that the time to the adjusted deadline is greater than the task execution time.

The scheduler maintains a queue for each worker according to the partitioned EDF scheme. To make a scheduling decision, a copy of each queue is scheduled using EDF and evaluated for potential deadline misses. Queues with predicted misses are eliminated. In addition, the overall density of the queue is calculated from the individual task \textit{densities}.

If the scheduler is configured to use the \textit{first-fit} strategy to select the worker, it will select the first worker without a missed deadline.
For the other strategies, each worker is tested.
In the case of the \textit{best-fit} or \textit{worst-fit} strategies, the scheduler selects the worker with the highest or lowest total density, respectively.
If the new task has the highest priority, the currently running task is preempted. 
The scheduler records the elapsed execution time $t_e$ of the task, and the worker saves its context.
If no worker is found for which its queue can be scheduled without predicted missed deadlines, the task is rejected.

\section{Evaluation}
\label{sec:evaluation}

\begin{figure*}
\centering
\includegraphics[width=1\linewidth]{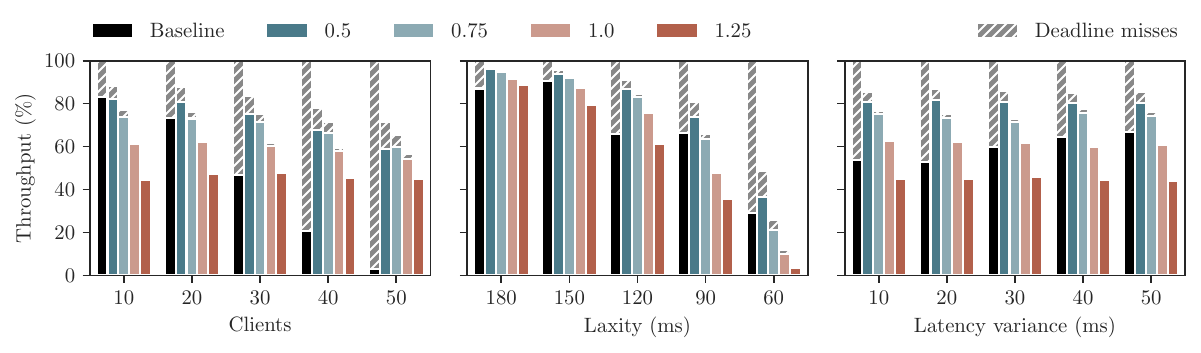}

\vspace{-6mm}\hspace{.05\linewidth}\subfloat[\label{fig:scalability-a} Scenario 1]{\hspace{.28\linewidth}}
\hspace{.03\linewidth}
\subfloat[\label{fig:scalability-b} Scenario 2]{\hspace{.28\linewidth}}
\hspace{.03\linewidth}
\subfloat[\label{fig:scalability-c} Scenario 3]{\hspace{.28\linewidth}}
\vspace{-2mm}
\caption{For each scenario, we report the throughput (successfully finished tasks) and the fraction of tasks that have been accepted but missed their deadline. We compare the reference baseline vs. our latency-aware approach at $\mathcal{U}=\{0.5, 0.75, 1.0, 1.25\}$.}
\label{fig:throughput}
\end{figure*}

We evaluate three scenarios to investigate the impact of our architecture on offloaded real-time tasks in wireless IIoT environments. 
We explore the limits at which tasks or network environments may or may not meet real-time requirements.

\subsection{Experimental Setup}
We implemented the proposed architecture and partitioned EDF scheduler in Rust and made the code publicly available\footnote{\url{https://github.com/dos-group/Real-Time-Offloading-Simulator-IIoT}}.
We implemented a reference scheduler for comparison. It runs EDF on a global task queue, rejecting tasks only if the time to deadline is less than the WCET and without deadline adjustment.

\paragraph{Network}
We use \textit{Mininet}\footnote{\url{https://mininet.org}} to emulate a network of clients, workers, and the scheduler.
The virtual network links between hosts can be configured with arbitrary latency, jitter, and bandwidth limits.
The basic network layout is the same for all test setups:
Client hosts are connected wirelessly to an access point, which is connected to a switch via Ethernet.
Edge cluster hosts (running workers and the scheduler) are also connected to this switch via Ethernet.

\paragraph{Test Scenarios}

\begin{table}
\centering
\caption{\label{tab:parameters} Test Scenario Parameters}
\begin{tabular}{l || c|c|c|}
\textbf{Scenario} &\textbf{1}&\textbf{2}&\textbf{3}\\
\hline
number of clients & 10 -- 50 & 30 & 30 \\
submission freq. mean [1/s] & 1 &1 &1\\
latency mean [ms] & 30&30&30\\
latency variance [ms] &10&10&10 -- 50 \\
laxity mean [ms] & 100 &180 -- 60&100\\
\end{tabular}
\end{table}

All tasks in a test run have the same predefined execution time of 30 seconds and have been run 5 times each.
The relative deadlines, on the other hand, are only configured in terms of mean and variance, and are dynamically sampled from a normal distribution by the clients at runtime.
This results in more or less preemptions, depending on the configuration, because the deadlines of the tasks differ accordingly for the same execution time.

The task submission frequency is sampled from a Poisson distribution for which $\lambda = 1$ is configured for all clients.
However, clients only submit one task at a time, so the minimum time between task submissions is limited by the task execution time.
The worst case task execution times are set to 100ms for all experiments. 

As described in section \ref{sec:adjustment}, the scheduler is configured with an uncertainty factor $\mathcal{U}$ and the task assignment heuristic.
The scheduler approximates the worker-to-client latency to be considered when creating the schedule based on the client-to-scheduler latency measured during task submission.
The measured client to worker latency is multiplied by $\mathcal{U}$ to account for variations in latency.
The following test scenarios explore how much variance needs to be accounted for with which uncertainty factor, and the impact this has on overall throughput.
Each scenario is tested with $\mathcal{U}$ ranging from 0.25 to 5.0. 
The table \ref{tab:parameters} lists the chosen parameters for the three scenarios.

\subsection{Scenario 1: Number of Clients}

The first scenario examines the behavior of the system under varying loads.
Ten to fifty clients each create an average of one task per second.

The different task assignment heuristics were compared to see if the tasks were distributed appropriately.
All reported results use the worst-fit allocation strategy, as first-fit and best-fit simply concentrate the workload on a single worker.

\paragraph{Results}

The experimental results can be seen in Figure~\ref{fig:scalability-a}. We present only the most relevant results for uncertainty factors between 0.5 and 1.25. For higher values, the acceptance rate drops without any further gain in reliability. We present the ratio of successfully offloaded tasks, where a submission is considered \emph{successful} if the scheduler accepts it and the computation result is returned before its deadline has passed. 

The results show that as the load increases, the reference miss rate increases, reducing the actual throughput of successful tasks. Depending on the uncertainty factor used, our approach minimizes the number of missed deadlines for any load. However, for higher loads and higher uncertainty factors, the number of rejected tasks increases, leading to an expected decrease in relative throughput. There are few to no missed deadlines in any of the experiments using our design and a success rate of up to 60\% for the highest tested submission load. Rejected tasks are returned to clients as early as possible for fallback processing.

\subsection{Scenario 2: Task Laxity}
Next, we test the effect of task laxity on the acceptance rate and the miss rate. The lower the laxity, the less communication and scheduling time is available. Experiments were conducted with laxities ranging from 180ms down to 60ms. 

\paragraph{Results}
It is evident that task laxity plays an important role in real-time processing. With sufficient laxity, the proposed design is able to meet all deadlines while rejecting only a small fraction of tasks, resulting in a relative throughput of 95\%, as can be seen in Figure~\ref{fig:scalability-b}. As the laxity becomes smaller, fewer tasks can be safely accepted. The proposed design outperforms the reference implementation for all laxities and for uncertainty factors below 1.0, meaning that some risk of missed deadlines must be accepted. The best results could be achieved with $\mathcal{U} = 0.5$, with a successful throughput difference of 5 - 20 percentage points compared to the reference scheduler. By choosing $\mathcal{U} > 1.0$, missed deadlines can be avoided for all tested laxities. However, in the most extreme case of 60ms submission laxity, only 5\% of submissions are accepted to accommodate this predictability. With $\mathcal{U} = 1.0$ and submission laxities above 90ms, more tasks are successful than with the reference scheduler.

\subsection{Scenario 3: Latency variance}
The final scenario tests the behavior of the system under increasing network latencies. We tested latency variances ranging from 10ms to 50ms.

\paragraph{Results}
The results for this scenario in Figure~\ref{fig:scalability-c} show the least impact of increasing the parameter. Again, we see that higher uncertainty factors lead to higher deadline hit rates and lower task acceptance rates. The relative throughput of our design outperforms the reference implementation with somewhat risky uncertainty factor values below 1.0.

\subsection{Summary}
\label{sec:eval_summary}

For very tight deadlines, where the laxity is in the same range as the latency, the real-time scheduler relies heavily on correct latency estimates to generate efficient schedules.
For low-variance latencies, it has been shown that high throughput can still be achieved while preventing missed deadlines and hence gaining overall system predictability.

For pessimistic latency estimates, stronger guarantees of meeting deadlines can be provided.
However, this comes at the expense of throughput, since a correspondingly large number of tasks must be discarded.
Still, high throughput can be achieved if isolated misses are acceptable. 
The choice of an acceptable uncertainty factor must be made taking into account client fallback mechanisms. 

The highest advantages of the proposed design can be seen under high task submission loads, where the reference design is only able to successfully compute 5\% of submitted tasks while missing the deadlines of the remaining 95\% of tasks. Here, our design has still a relative throughput of 45\% with no deadline misses. 

\section{Conclusion}
\label{sec:conclusion}
In this work, we investigated the trade-offs and implications of offloading real-time tasks over wireless networks to local edge resources in the context of IIoT environments.
On this basis, a system architecture was designed that integrates networking uncertainties.
This system was implemented and extensively tested.
\balance
Given the broadness of the proposed architecture optimizations and adaptations towards specialized use-cases can be applied to the scheduler and real-time networking concepts can be incorporated in future work. Scheduling algorithms could incorporate network bandwidth, memory consumption, and payload sizes. The distributed architecture will additionally be extended by fault tolerance mechanisms and removing the single point of failure.

\begin{acks}
    This research was supported  by the German Ministry for Education and Research (BMBF) as Software Campus (grant 01IS17050).
\end{acks}

\bibliographystyle{ACM-Reference-Format}
\bibliography{bibliography}

\end{document}